\documentstyle[11pt,newpasp,twoside,epsf]{article}
\begin{document}
\title{Quasi-Periodic Changes in the Parsec-Scale Jet of 3C\,345}
\author{J. Klare, J.A. Zensus, A.P. Lobanov, E. Ros, T.P. Krichbaum, A. Witzel}
\affil{Max-Planck Institut f\"ur Radioastronomie, Auf dem H\"ugel 69, D-53121 Bonn, Germany}

\begin{abstract}
We have observed the QSO 3C\,345 with VSOP at 1.6\,GHz and 5\,GHz and with the VLBA at 
22\,GHz and 43\,GHz, with 24 epochs in total, from 1997.40 to 1999.69. This study, combined 
with previous observations, reveals periodic changes in the parsec-scale jet.
This quasi-periodicity can be found in the initial position angles, the trajectories and the 
flux density evolution of the jet components, with a period of about 9\,yr. This period should 
reflect the dynamics of the material in the central region of 3C\,345 and it is possibly related 
to a putative binary black hole in the center of this object.
\end{abstract}

\section{Introduction}
The 16th magnitude quasar 3C\,345 ($z=0.595$) can be regarded as one of the archetypical sources
for studies of superluminal motion with components moving along strongly curved trajectories
on pc-scales. Several components with apparent velocities of $2-20c$ have been observed in the jet of 3C\,345
(Zensus, Cohen, \& Unwin 1995; Lobanov 1996). Periodic outbursts occurring every $10.1 \pm 0.8$\,yr 
were found by Zhang, Xie, \& Bal (1997) from the B-band historical light curve.

The ejection directions of the jet components from the core (the optically thick region) vary, and the component 
trajectories differ significantly from each other (e.g. Zensus et al. 1995). The curvature of the component trajectories 
increases towards the core, while the trajectories straighten at larger core separations.
This makes the innermost region the most interesting for studying the kinematics of the jet components. 

\section{Observations}
Seven VLBA observations at 22\,GHz and nine VLBA observations at 43\,GHz were made between 1997.40 and 1999.64. 
Additionally, four space-VLBI (VSOP) observations were performed at 1.6\,GHz and four at 5\,GHz between 
1998.22 and 1999.69. We also reanalysed 22\,GHz data from Ros, Zensus, \& Lobanov (2000) at epochs 
1996.41 and 1996.82 and 43\,GHz data from A. Marscher in epochs 1998.15, 1998,41, 1998.58, 1998.76, and 1998.94.
Altogether the data base contains 14 epochs at 43\,GHz,
9 epochs at 22\,GHz, 4 epochs at 5\,GHz and 4 epochs at 1.6\,GHz. A detailed
description of this work can be found in Klare (2003).

The observations allowed us to study in detail 5 jet components (C7--C11) in the inner 3\,mas. Three of them 
were ejected during the time period covered by the new observations 
and could be followed from the beginning of their evolution, much more closely  
than it was done for older components. To investigate the long-term behavior of the jet, 
the new observations were complemented by all data available in the literature, covering the period
from 1979 to 2000.

Model fitting by Gaussian components was performed in DIFMAP to parameterise the jet structure and 
study its evolution. The core of the jet is used as the stationary reference point, and 
the offset of the jet components is measured with respect to it. The observed proper motion of a component in right 
ascension and declination is fit by polynomials. To improve the fits, we determined the core shift
with respect to compact optically thin jet features, allowing us to combine observations 
at different frequencies. This procedure improved significantly the accuracy of the trajectory determination
for the younger components C7, C8 and C9. This led to a much better determination of their physical jet parameters.

\section{Similarity and quasi-periodic changes in the jet}
\label{quasiperiod}
There is a clear trend of the trajectories of the jet components C4 to C7 in 3C\,345 as shown
in Figure 1. C6 was a peculiar faint component and disappeared at about 1\,mas core distance. The new
data for C8 show for the first time a similarity in shape and curvature of the trajectories between two 
different components, namely C5 and C8. The younger jet component C8 passed through similar stages such as 
the turning point in its trajectory 8--10\,years later than C5.

\begin{figure}
\plotfiddle{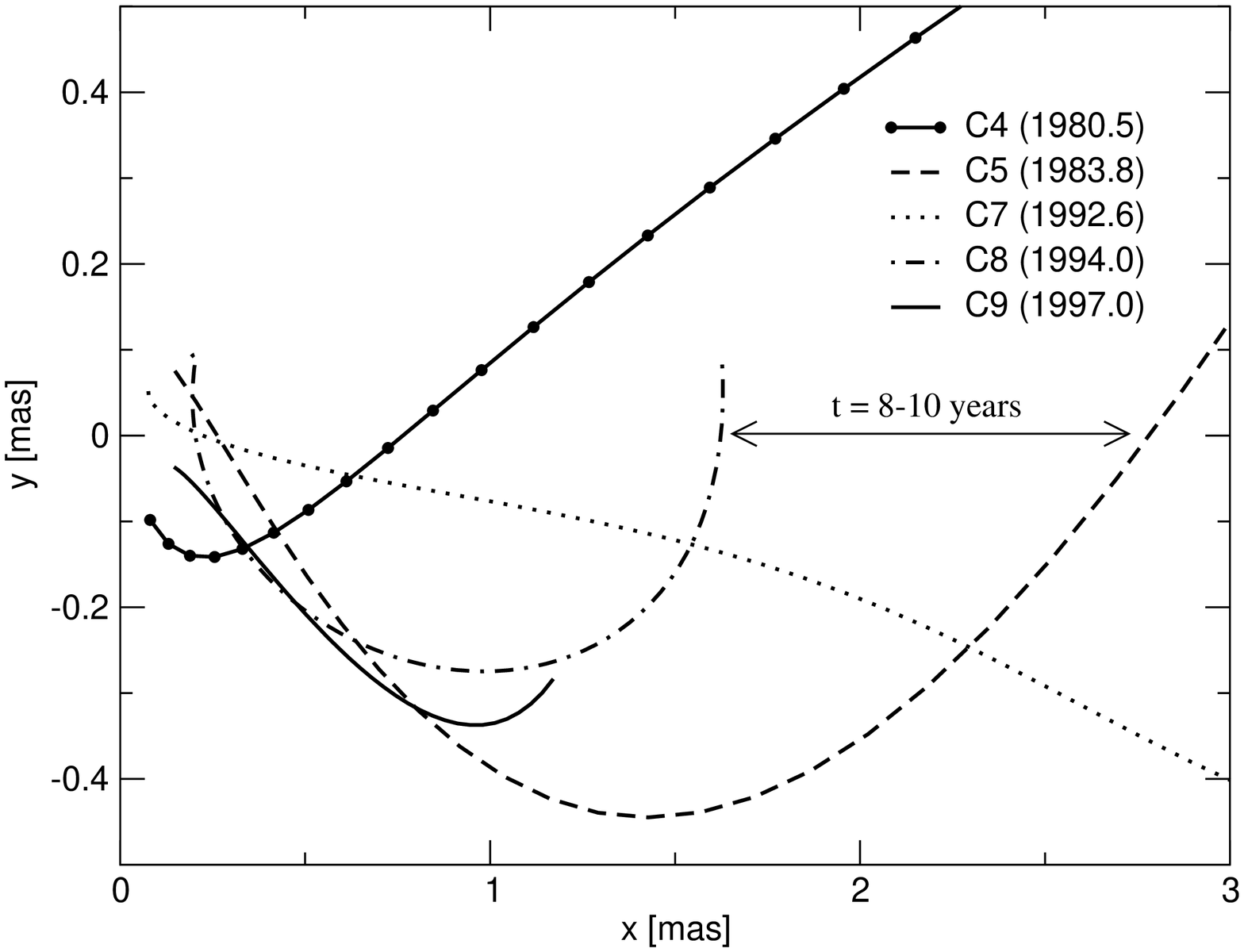}{80mm}{0}{48}{48}{-180}{-37}
\caption{Polynomial fits to jet component trajectories for C4, C5, C7, C8 and C9 in 3C\,345. 
Remarkable is the similar shape of the trajectories C5 and C8. The numbers in brackets are epochs at core 
distances of about 0.25\,mas.\label{C5C8C9}}
\end{figure}

Polynomial fits to jet component trajectories for C4, C5, C7, C8 and C9 in 3C\,345. Remarkable is the similar
shape of the trajectories C5 and C8. The numbers in brackets are epochs at core distances of about 
0.25\,mas.\label{C5C8C9}

The direction of the component ejection (P.A. near the core) changes regularly, with a period of 
\mbox{8--10\,years} (shown by the dashed line in Figure 2). Additionally, there is an overall trend 
(shown by the two dot-dashed lines) of $2.6^\circ \mbox{year}^{-1}$.

\begin{figure}
\plotfiddle{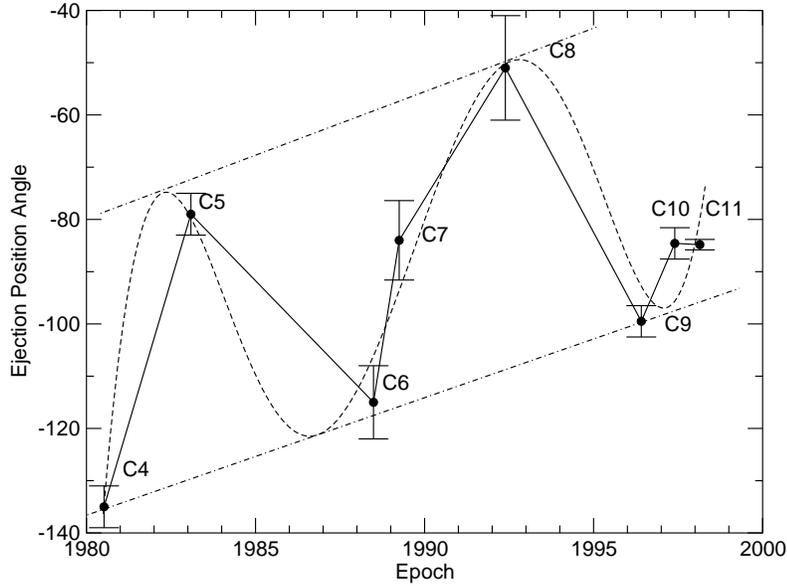}{80mm}{0}{46}{46}{-180}{-33}
\caption{Position angle of the ejection point of the different components of 3C\,345 since 1980.
A periodicity of 8--10\,years is visible (dashed line). The dot-dashed lines show an overall change
of the initial P.A. of $2.6^\circ \mbox{yr}^{-1}$.\label{paver}}
\end{figure}

The maxima in flux density of the different jet components as a function of time are shown 
in Figure 3. A 9.0\,yr period is observed here. Comparison of the jet trajectories
with the component flux density maxima reveals a correlation. The turning-points in the trajectories
coincide with the component flux density maxima. Furthermore, a detailed kinematical analysis of the component 
trajectories show that the most plausible scenario is that of an increasing Lorentz factor along the trajectories
(Klare 2003; Lobanov \& Zensus 1996, 1999). 
From these follows, that the viewing angles of the components decrease shortly after ejection to $\sim 1 ^\circ$ 
and turn back to higher values. The component flux density peaks are thus caused by Doppler boosting.
This means, that a periodic geometrical change of the jet causes the observed 9.0\,yr period in the
component flux density maxima. 

\begin{figure}
\plotfiddle{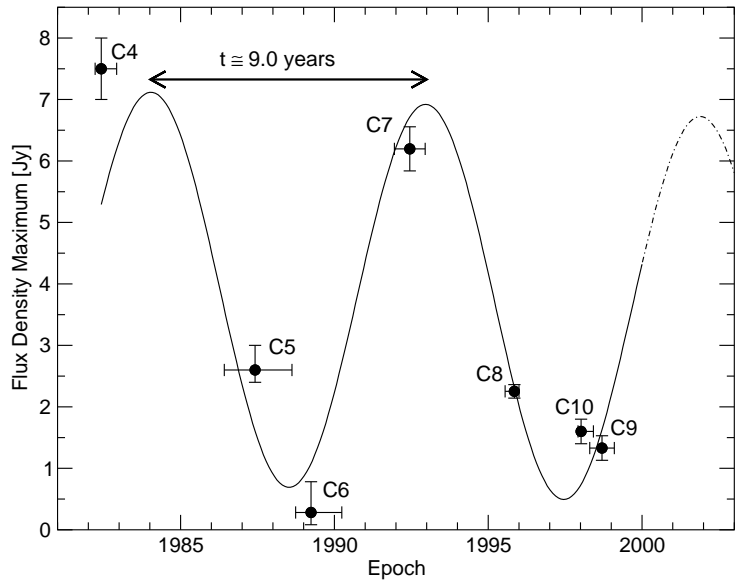}{8cm}{0}{46}{46}{-180}{-33} \\
\caption{Flux density maxima for the jet components in 3C\,345 at 22\,GHz as a function of time. A sinusoidal
function with a 9.0\,yr period can be fitted.\label{fluxrelepoch}}
\end{figure}  

\section{Discussion}
The results presented above emphasize the existence of an underlying quasi-periodic process 
in the pc-scale jet of 3C\,345 with a time period of about 9\,yr. A possible scenario, which can explain the 
observed quasi-periodicity, is the binary black hole (BBH) system (e.g., Begelman, Blandford, \& Rees 1980).
The BBH model was applied to describe the optical light curve and the kinematic and flux density evolution
of the jet component C7 in 3C\,345 (Lobanov \& Roland 2002). The model yields an orbital period of 170\,yr and
a precession period of 2500\,yr. The $2.6^\circ \mbox{year}^{-1}$ of the initial P.A. shown in Figure 2 may
be related either to the orbital or to the precession period in this system. The observed 9\,yr period is too short to
be related to
the orbital period in the BBH system. The 9\,yr period is more likely to be caused by the rotation of the 
jet or the rotation of the accretion disc. Further work has to be done to determine the origin of the 9\,yr period. 

\begin{acknowledgements}
The National Radio Astronomy Observatory is a facility of the National Science Foundation operated under cooperative agreement by 
Associated Universities, Inc. We gratefully acknowledge the VSOP Project, which is led by the Japanese Institute of Space and 
Astronautical Science in cooperation with many organizations and radio telescopes around the world. We thank A. Marscher 
for sharing with us the VLBA data at 43\,GHz in 1998.

\end{acknowledgements}

\end{document}